\documentclass{article}
\usepackage{spconf,amsmath,amssymb,graphicx,bbm,multicol,subfig}
\ninept
\DeclareUnicodeCharacter{200E}{}
\title{TOWARDS DATA-EFFICIENT MODELING FOR WAKE WORD SPOTTING}
\name{Yixin Gao, Yuriy Mishchenko, Anish Shah, Spyros Matsoukas, Shiv Vitaladevuni
}
\address{Alexa, Amazon.com Services LLC\\
101 Main St, Cambridge, MA 02142, USA}
\begin{document}
\ninept
\maketitle
\begin{abstract}
Wake word (WW) spotting is challenging in far-field not only because of the interference in signal transmission but also the complexity in acoustic environments. Traditional WW model training requires large amount of in-domain WW-specific data with substantial human annotations therefore it is hard to build WW models without such data. In this paper we present data-efficient solutions to address the challenges in WW modeling, such as domain-mismatch, noisy conditions, limited annotation, etc. Our proposed system is composed of a multi-condition training pipeline with a stratified data augmentation, which improves the model robustness to a variety of predefined acoustic conditions, together with a semi-supervised learning pipeline to accurately extract the WW and confusable examples from untranscribed speech corpus. Starting from only 10 hours of domain-mismatched WW audio, we are able to enlarge and enrich the training dataset by 20-100 times to capture the acoustic complexity. Our experiments on real user data show that the proposed solutions can achieve comparable performance of a production-grade model by saving 97\% of the amount of WW-specific data collection and 86\% of the bandwidth for annotation. 
\end{abstract}
\vspace{-0.5em}
\section{Introduction}
\label{sec:intro}
% WW problem
Wake word (WW) is the gatekeeper that enables end users to interact with the cloud-based voice assistants. To ensure good customer experience, a WW spotter needs to be highly sensitive to detect the WW from low signal-to-noise ratio (SNR) speech, and specific to minimize false alarms, thereby protecting users' privacy. WW spotters are widely deployed on the voice-enabled smart home devices such as Amazon Echo, Google Home, Apple HomePod, etc. These devices are often installed in complicated acoustic conditions and used from a distance, therefore the speech signal that has arrived to a device is heavily interfered and attenuated by the dynamical environment, resulting in many difficulties in far-field WW detection. 

% literature in KWS
Tremendous efforts have been made in recent years for improving far-field keyword spotting or WW detection. Most of them focused on investigating novel model architectures\cite{Chen2014_DNN, Sainath2015_CNN, He2017_Seq2Seq, Sun2017_TDNN,Kumatani2017_RAW,Shan2018_Attn,Guo2018_Highway, Wu2018_Monophone} and improving the efficiency of training\cite{ Panchi2016_Multitask,Tucker2016_SVD,Raju2018_Augm,Li2018_TS, Chen2018_MAML}. These efforts implicitly assume that the transcribed in-domain WW-specific training data is sufficient for building WW models. The training data used in these works ranged from 200 hours to tens of thousands of hours, and most of them were collected from far-field devices in real use and transcribed by human annotators to ensure that the richness of the background is accurately captured by the training data. 

% summary of this paper
For a new WW, collecting and transcribing such large size of speech corpus from matched domain, for example far-field,  is time-consuming and cost-prohibitive, which can be a roadblock for building a spotter for the new WW. Therefore how to develop data-efficient solutions to WW model building becomes the key question. In this paper, we develope solutions to build far-field WW spotter using as little as 10 hrs ($\sim$ 10K utterances) of the WW training data that was collected from close-talk microphones (CTM) in a quiet environment.  In many situations CTM data is readily available, and compared with far-field data is much easier to obtain, for example by using croud-sourcing \cite{speechcmd2018}. Our approach uses a stratified data augmentation for multi-condition training, together with a semi-supervised learning framework employing an automatic speech recognition (ASR) model to extract WW and confusable examples from untranscribed speech corpus. 

In literature data augmentation has been widely used in ASR  \cite{Ko2015_Augmentation,Park2019_Augmentation,Ko2017_Augmentation,Raju2018_Augm}, however traditional approaches simplistically apply arbitrary distortions, or random clipping, or transformation of audio in single pass indiscriminately. Such strategy is much less effective when working with extremely limited amount of data. On the contrary, we first define specific use cases, such as clean close-talk, noisy close-talk, clean far-field, and noisy far-field, and further explore how these cases, generated from limited number of initial clean utterances, should be used in the training dataset for the best performance of WW spotting in real world. 

%Compared with the efforts to use untranscribed speech data via unsupervised learning and semi-supervised learning in tasks such as ASR and spoken term detection (STD) \cite{Li2018_TS, Manohar2018, Yeh2019, Hari2019_MHr, Wu2019_TS}, the amount of transcribed training data required for those tasks is much greater than in this paper. 
There are many open questions for using untranscribed data for WW spotting, for example, how to accurately extract WW data from inaccurate results of ASR recognition; how to balance the mined dataset to avoid bias; how to automatically scale up to arbitrary new WW, new language, etc. For discriminative training, how to define the hard examples and how to efficiently obtain them is unknown. Traditional approach of blindly sampling from non-WW audio \cite{Chen2014_DNN, Sainath2015_CNN,Panchi2016_Multitask} does not optimize the model discrimination. In this paper, we answer these questions by the proposed semi-supervised learning framework equipped with a mechanism called ``lexicon filter" to effectively extract discriminative examples for a given WW.

Our experiments on real user data showed that the proposed methods enable us to build WW models that matches the performance of a production-grade WW models trained with several hundreds of hours of transcribed WW data collected from real far-field scenarios. To the best of our knowledge, it is the first work in literature that is focused on how to use such little training data for building production-quality far-field wake word spotters.  In addition, we benchmark the performance gain from each method to provide practical references for a variety of use cases for building new WW spotters  such as CTM data only; clean far-field data only; untranscribed speech corpus only, etc

\vspace{-0.5em}
\section{System overview}
\label{sec:system}
%system
\vspace{-0.5em}
\begin{figure}[t]
\centering
\includegraphics[width=0.5\textwidth]{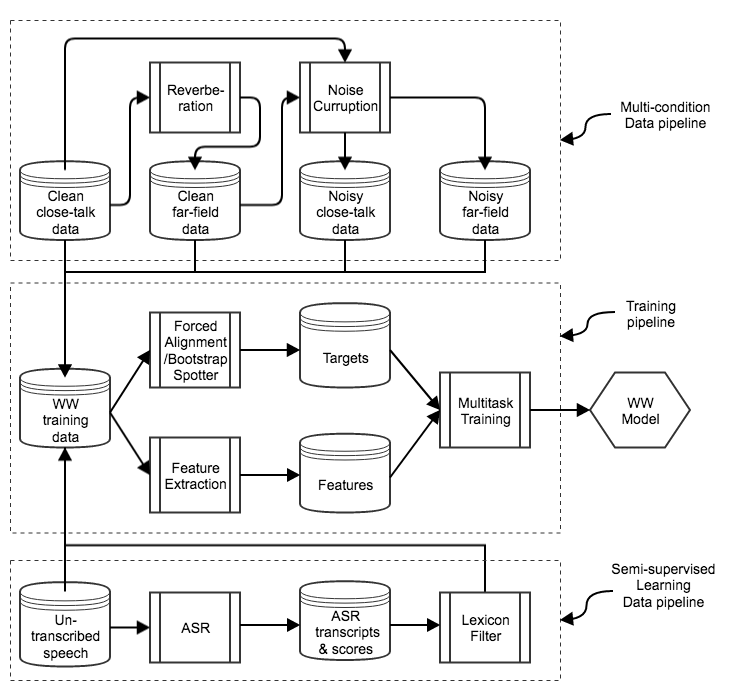}
\caption{The proposed WW training system.}
\label{fig:system}
\end{figure}
\vspace{-0.5em}
%We start by developing a data augmentation strategy for multi-condition training to improve the WW model robustness to various environments that are weakly represented by the original CTM data. The CTM data was first convolved with audio recordings from a library of room impulse response (RIR) and then corrupted by additive noise profiles at different levels of SNR to simulate the environmental conditions of clean near-field, noisy near-field, clean far-field and noisy far-field environment. Next, we introduce a semi-supervised learning (SSL) strategy to utilize the untranscribed far-field speech for better modeling the complex acoustics and improving the discrimination of confusing cases. Our experiments show that by employing the proposed augmentation and learning strategies we are able to match the performance of a production-grade WW models trained with several hundreds of hours of WW data collected from real far-field scenarios. In addition, we benchmark the performance gain from each strategy to provide practical references for a variety of use cases for building new keyword spotting models such as CTM data only; clean far-field data only; untranscribed speech corpus only, etc.  
The proposed training system for WW spotter is shown in the Figure \ref{fig:system}. The system is composed of two data pipelines: multi-condition and semi-supervised learning, and a multi-task training pipeline. 

The multi-condition data pipeline applies stratified data augmentation to a small set of clean close-talk speech data. Specifically, the CTM data was first convolved with audio recordings from a library of room impulse responses (RIR) to mimic the reverberated speech received by far-field devices. Then both the CTM and reverberated speech were corrupted by additive noise profiles at different levels of SNR, generalizing the CTM data to become a representation of the audio signal in typical household use cases, i.e. mixed conditions of clean near-field, noisy near-field, clean far-field and noisy far-field environments. %to simulate speech data collected from mixed  environmental conditions of clean near-field, noisy near-field, clean far-field and noisy far-field environment with various background noises. 

The semi-supervised learning data pipeline uses an ASR model to obtain the automatic transcripts with confidence scores, and is equipped with a mechanism called ``lexicon filter" to mine utterances containing WW and confusable words from untranscribed speech corpus. The Lexicon filter uses the Levenshtein distance as a measurement of pronunciation proximity to filter automatic transcripts. Each data pipeline can be plug into the training system separately, and they can be combined together to complement each other. We will discuss the effects of each method in section \ref{sec:expr}.

\vspace{-0.5em}
\section{Multi-condition training}
\label{sec:augm}
%augmentation
Multi-condition training (MCT) \cite{Seltzer2013_MCT}, i.e., training with data from heterogeneous sources and diversified conditions, enables a model to learn higher level features that are more invariant to noise, and was shown to be effective in robust ASR \cite{Huang2014_MCT,Gibson2018_MCT} . In this paper we use data augmentation to simulate training data associated with conditions that are weakly represented in the original CTM data, such as far-field and/or noisy environment, and employ the mixed-condition training approach  \cite{Huang2014_MCT} to train a DNN WW spotter. Our data augmentation strategy for multi-condition training contains two components: reverberation and noise corruption. %Reverberation transforms the CTM speech via convolution with room-impulse response (RIR) filters to mimic speech received by a device in far-field. Corruption of the training data is achieved by artificially mixing it with different noise samples of background household noise and music playback at various signal-to-noise ratios, generalizing it to become a representation of the audio in typical real world household use cases.
\subsection{Reverberation: from close-talk to far-field}
\label{sec:rir}
The far-field speech can be obtained by convolving the close-talk speech with room-impulse response filters:
\vspace{-0.5em}
\begin{equation}
\label{eq:reverb}
x_R(t) = x_S(t)\ast r(t)
\end{equation} 
where $r$ is room impulse response,  $x_S$ is source close-talk signal and $x_R$ is the reverberated signal, respectively. The image-source method, as presented originally in \cite{Allen1979_ImageMF}, can be used to precisely simulate the RIRs for small rooms with known measurements. However the computation grows exponentially with the order of reflections thus preventing it from practical usage.% in distant and complex environment.

In this paper we applied a proprietary library of RIR filters \cite{Raju2018_Augm} to mimic far-field audio. The RIR filters library was collected through real devices set up in real rooms simulating household living rooms by playing audio signal such as chirps at different locations.  By varying the locations of the devices or the sound sources, the RIR library approximated the effects of playing the audio at random locations in the room.% The library can be used not only for WW modeling, but other far-field systems and applications such as ASR, acoustic event detection (AED), speaker identification (SID), etc. 

%When applying the reverberation to each utterance in the CTM dataset, a random filter is chosen from the list of RIR filters in the library, and convolved with the CTM audio to generate a far-field version of the utterance. We call such a traversal of the CTM dataset a pass. In each pass we randomly select a different RIR filter for each given utterance. Since they are all unique we can multiply the amount of the simulated far-field data by running the reverberation passes multiple times.% on the same CTM dataset. Each pass mimics a far-field data set with a certain instance of device placements.
%
\subsection{Corruption: improving the noise robustness}
\label{sec:corruption}
Since our close-talk data is from clean acoustic environment, the reverberated data is also acoustically clean, hence it does not reflect well the real noisy conditions. %A WW spotter trained with such clean data does not generalize well into noisy conditions. 
We employ two internally collected noise data libraries: household noise and music/movie audio, to improve the performance in both general household and media playing conditions. The simulated noisy far-field data can be achieved by %Since our target use cases are far-field, we first reverberated the music/movie audio using \eqref{eq:reverb}, then mix the original or reverberated data according to \eqref{eq:corrup}:
\vspace{-0.5em}
\begin{equation}
\label{eq:corrup}
x_A(t) = x(t) + \alpha\cdot n(t) + \beta\cdot m(t)
\end{equation}
where $x\in x_R\cup x_S$ is the clean original CTM or reverberated signal, $x_A$ denotes the corrupted signal, $n$ and $m$ are noise or music interferences, respectively. $\alpha$ and $\beta$ are the scaling factors to control the ratio of the two noise sources and they are coupled with the target SNR, which is randomly drawn from a normal distribution with mean 10.0 dB and standard-deviation 3.0 dB. %Similar to the reverberation step, a traversal of corruptions to a dataset is called a pass, and we can multiply the amount of data by repeating the noise corruption pass.
\subsection{Mixed-condition training}
We use a combination of the above data augmentation methods to create a mixed-condition training set of clean, clean+reverb, clean+noise, clean+reverb+noise speech. %, simulating the speech from multiple conditions including quiet close-talk, noisy close-talk, clean far-field and noisy far-field with various SNR.  
Any labels, such as the DNN targets,  of the augmented data are the same as the labels of original data. We pool the mixed-condition data together and train a single DNN model, which is also called mixed-conditional model or multi-style model \cite{Huang2014_MCT}.%, and show that such model demonstrates robustness to a diverse range of use cases. 
%Figure \ref{fig:multi-condition} shows the waveform of the simulated multi-condition data for the same WW utterance.  
%
%\begin{figure}[htp]
%\includegraphics[width=0.45\textwidth]{figures/multi_condition}
%\caption{Waveform of a WW from multiple conditions (topleft, topright, bottomleft, bottomright): clean, clean+reverb, clean+noise, clean+noise+reverb. }
%\label{fig:multi-condition}
%\end{figure}

%
\vspace{-0.5em}
\section{Semi-supervised learning}
\label{sec:ssl}
%data mining from untranscribed far-field speech
%general speech is easier to aquire, however
%ssl
%Our data augmentation strategy simulates the acoustic conditions occuring during signal transmission. However 
The augmented dataset is limited by the diversity of the source, such as the size of the vocabulary, the population of speakers etc.  A general speech corpus used for training ASR models, which usually contains thousands of hours of audio and human-prepared parallel transcripts, incorporates much greater variety in sources, nevertheless the majority of such speech corpus is not relevant to WW. There may exist large corpora of untranscribed audio data with significant amounts of WW-relevant data, but they are hard to use because of lack of human annotations. %For WW modeling, an effective usage of untranscribed speech data therefore becomes critical. 

In this paper we develop a simple but efficient semi-supervised learning (SSL) strategy to mine useful WW-training data from large untranscribed general-speech corpora. %Compared with the efforts to use untranscribed speech data via unsupervised learning and semi-supervised learning in tasks such as ASR and spoken term detection (STD) \cite{Li2018_TS, Manohar2018, Yeh2019, Hari2019_MHr, Wu2019_TS}, the amount of transcribed training data required for those tasks is much greater than in this paper. 
We employ a general far-field ASR model to automatically transcribe the unlabeled speech corpus, then select WW-specific utterances and add them to our training set. % and produce frame-wise targets for DNN training. 
%First we run the ASR decoder on the unlabeled speech data to generate the automatic transcripts. Next, we select utterances containing the WW and add them to our training set. 
The DNN targets are obtained by force aligning the speech to the automatic transcripts. Since the ASR transcript may not be accurate,  only the utterances with WW confidence score higher than a threshold $\theta_P$  are selected. It should be noted that, in the utterances mined here the WW may appear in the middle of the utterance, therefore the phonetic context may be different from the utterances initiated by the WW. In practice, our experiments show that such context mismatch have unobservable effects on WW spotting. 
\vspace{-0.5em}
\subsection{Lexicon filter}
\label{sec:confusable}
\begin{figure}[tp]
\centering
\includegraphics[width=0.42\textwidth]{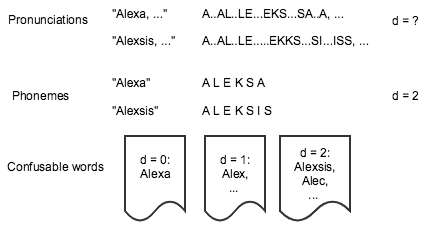}
\caption{Lexicon filter for ``Alexa".}
\label{fig:alexa}
\end{figure}

In machine learning overemphasizing hard examples during training commonly leads to better model discrimination and generalization. %For a particular WW, adding the confusable examples to the training data can improve the robustness of the WW model. 
For a particular WW, ideally a hard example %could be a word, a subword, a phrase, or any 
is a combination of syllables that pronounces similarly to the WW. However accurately calculating the similarity of pronunciations of words in two sentences is computationally intractable. Here we propose a mechanism called ``lexicon filter", which calculates the Levenshtein distances \cite{Navarro2001} with the target WW over the phoneme representations of words provided by the lexicon from the ASR model as an approximation of the pronunciation similarity to define the set of confusable words for the WW. Since the word dictionary in ASR model is usually very large, and the majority of the words rarely appear in a speech corpus, constraining the search within the most frequent words would save a lot of computations. This set is used to filter the ASR transcripts for utterances containing the confusable words. Utterances containing such words with confidence score higher than a threshold $\theta_N$ are added to training set as the negative examples. Without loss of generality we use the popular wake word ``Alexa" to explain the concept of the lexicon filter in Fig. \ref{fig:alexa}.
\vspace{-0.5em}
\subsection{Loss function}
The SSL loss for DNN training is
\vspace{-0.5em}
\begin{equation}
\begin{aligned}
L_{SSL} =\sum_{\substack{x \in \mathbf{u} \\ \mathbf{u}\in P\cup N}}
\left(\mathbbm{1} (\mathbf{u}\in P) y\log \frac{1}{q(x)}\right.
+\left.(1-y)\log\frac{1}{1-q(x)}\right)  
\end{aligned}
\end{equation}
where $P$ denotes the set of WW (positive) examples ($d = 0$) and $N$ denotes the set of confusable (negative) examples ($d \geq 1$), $x\in \mathbf{u}$ denotes the context-dependent feature frame in an utterance $\mathbf{u}$, $q(x)$ denotes the DNN output posterior probability for $x$ , and  $y \in \mathbf{y}$ is the corresponding framewise DNN target obtained by forced alignment using the ASR transcripts, i.e.
\vspace{-0.5em}
\begin{equation}
\mathbf{y} = fa(\mathbf{u}, ASR(\mathbf{u}))
\end{equation}
where $\mathbf{u}$ denote the utterances in the untranscribed speech corpus, $fa$ denotes the forced alignment operator\cite{Jelinek76_ASR}, respectively.  

The thresholds $\theta_P$ and $\theta_N$ can be tuned to balance the ratio of positive and negative examples in the training set. In our experiments  we choose both $\theta_P$ and $\theta_N$  as $0.5$. In addition,  for a six-phoneme wake word, we found the lexicon filter threshold on the Levenshtein distance $d = 1, 2$ produces sufficient negative examples for augmenting the training with confusable cases.

\vspace{-0.5em}
\section{Experiments}
\label{sec:expr}
%experiments
\vspace{-0.5em}
\subsection{Model architecture}
\begin{figure}[htp]
\centering
\includegraphics[width=0.48\textwidth]{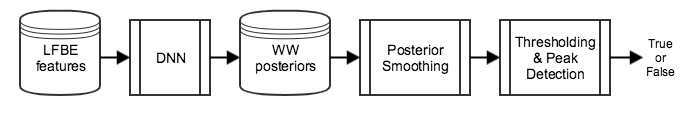}
\caption{The WW spotter.}
\label{fig:mtdnn}
\end{figure}

% dnn spotter
The WW spotting system used in this paper is composed of a compressed feed-forward deep neural network (DNN),  a posterior smoothing and a peak detection modules, as shown in Fig. \ref{fig:mtdnn}.  The input audio signal to the WW spotter is resampled at 16KHz and a 20-bin log filterbank energy ( LFBE) feature is computed every 10 msec with a 25 msec observation window. The LFBE features from a context window of 31 frames (20  on the left and 10 on the right) are stacked into a 620-dimensional vector and fed into the DNN. 
% The DNN maps the audio feature to the posterior probabilities of the input frame being a wake word or background. 
The DNN is composed of three non-linear fully-connected layers with 400 nodes, and before each non-linear layer a linear bottleneck layer with 87 nodes is inserted by decomposing the weighting matrix into two smaller matrices, similar to \cite{Tucker2016_SVD,Sun2017_TDNN,Sainath2013_SVD}. The output layer is a two-dimensional softmax, which can be easily extend to multiple WWs. For decoding, the posteriors corresponding to the WW are first smoothed by a windowed moving average, where the window size is approximately the same as the average duration of the WW, then a thresholding and peak-detection algorithm is applied to infer both the location and duration of a WW instance.
%by applying a 20-bin Mel-filterbank on the power spectrum of the audio signal and computing the the logarithm of the sum of the energy from each bin.
\vspace{-0.5em}
\vspace{-1em}
%
%for example a 400x400 matrix is decomposed into two matrices of sizes 400x87 and 87x400,
%as shown in Fig. \ref{fig:arch}. The introduction of the linear bottleneck layers play as low-rank approximation of the weight matrix between two successive non-linear layers and has effectively compressed the size of the DNN \cite{Tucker2016_SVD,Sun2017_TDNN}. For our case, compared with the equivalent fully-connected three-layers non-linear DNN, which has 570K parameters, the the total number of parameters of the compressed DNN is approximately 230K. 
%\begin{figure}[tp]
%\centering
%\includegraphics[width=0.4\textwidth]{figures/arch}
%\caption{The DNN architecture of the WW spotter.}
%\label{fig:arch}
%\end{figure}  
%
%The output layer of the DNN is a two-dimensional softmax, which produces the posterior probabilities of the input audio feature belonging to the phoneme set of the WW or the background. This architecture extends straightforwardly to the case of multiple wake words by modifying the dimensions of the output softmax layer. For decoding, the posteriors corresponding to the WW dimension are first smoothed by a windowed moving average, where the window size is approximately the same as the average duration of the WW, then a thresholding and peak detection algorithm is applied on the smoothed posteriors to find out the peak that exceeds a certain threshold. The location and the duration of the WW can be inferred from the location of the peak and the threshold-crossing of the smoothed posteriors.
\vspace{-0.5em}
\subsection{Data sets}
We choose ``Amazon" as our new wake word for experiments. A total of 9487 utterances beginning with the chosen WW were recorded from a close-talk microphone in quiet room, denoted as "CTM10K" dataset. The average duration of the utterances is 3.86 secs and the total duration of the dataset is 10.18 hrs.  For comparison, a baseline production-grade WW model was trained using 375 hours of transcribed speech data collected from far-field devices in real use cases.

The evaluation dataset contains a composition of more than 300 hours of audio collected from far-field, mid-field and near-field devices in real use cases from various acoustic conditions. The dataset is sufficiently large to show strong statistical significance.
\begin{figure*}[htp]
\centering
\subfloat[][]{
\includegraphics[width=0.67\columnwidth]{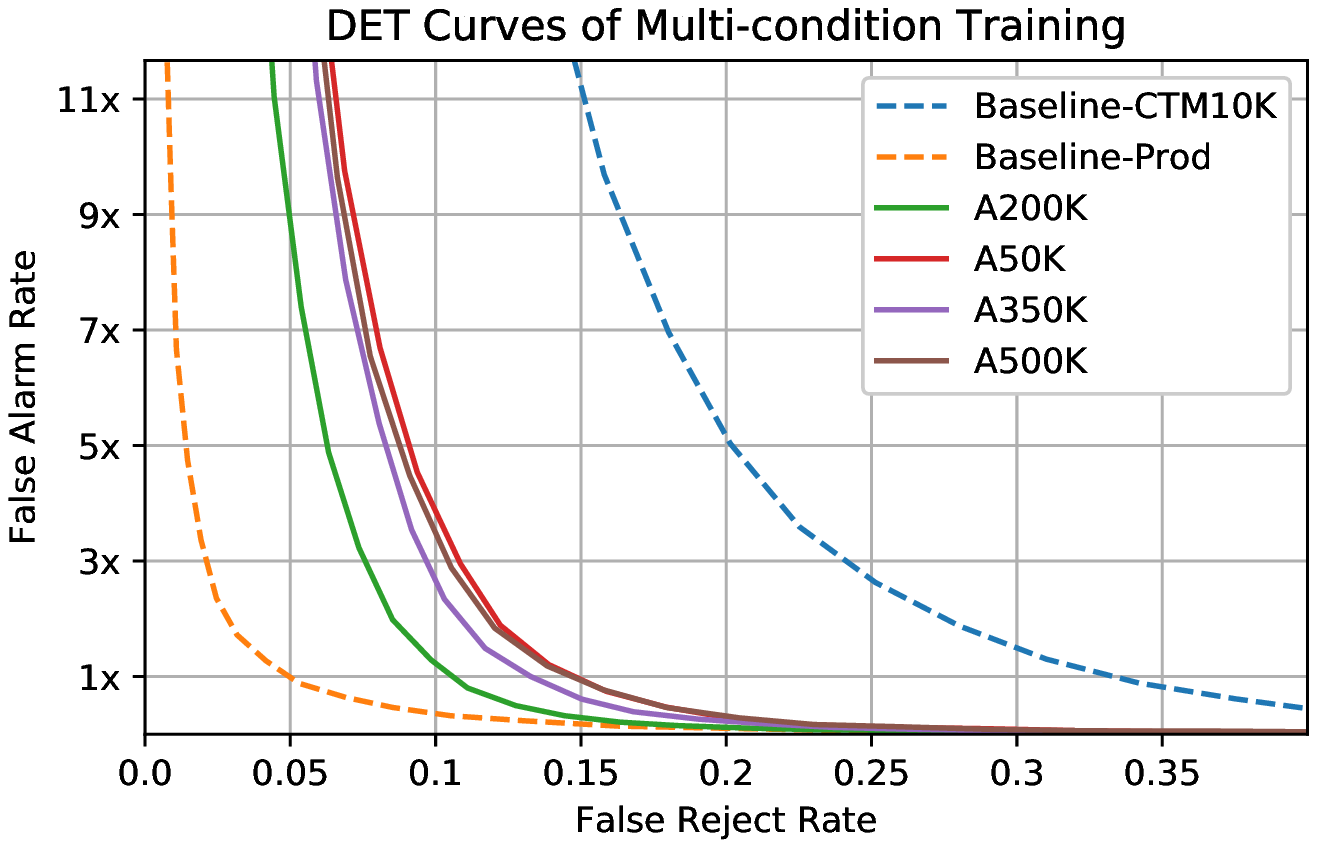}}
\subfloat[][]{
\includegraphics[width=0.67\columnwidth]{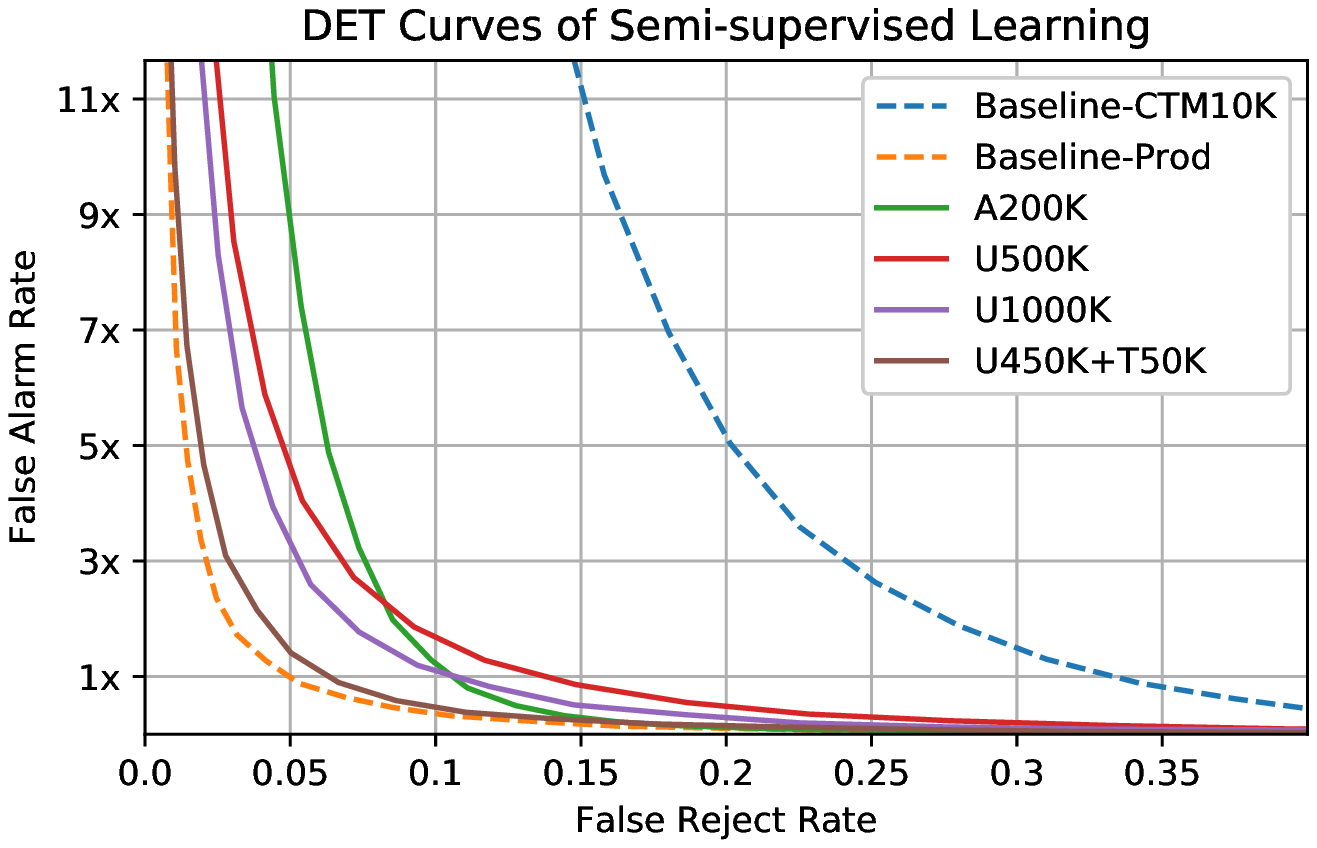}}
\subfloat[][]{
\includegraphics[width=0.67\columnwidth]{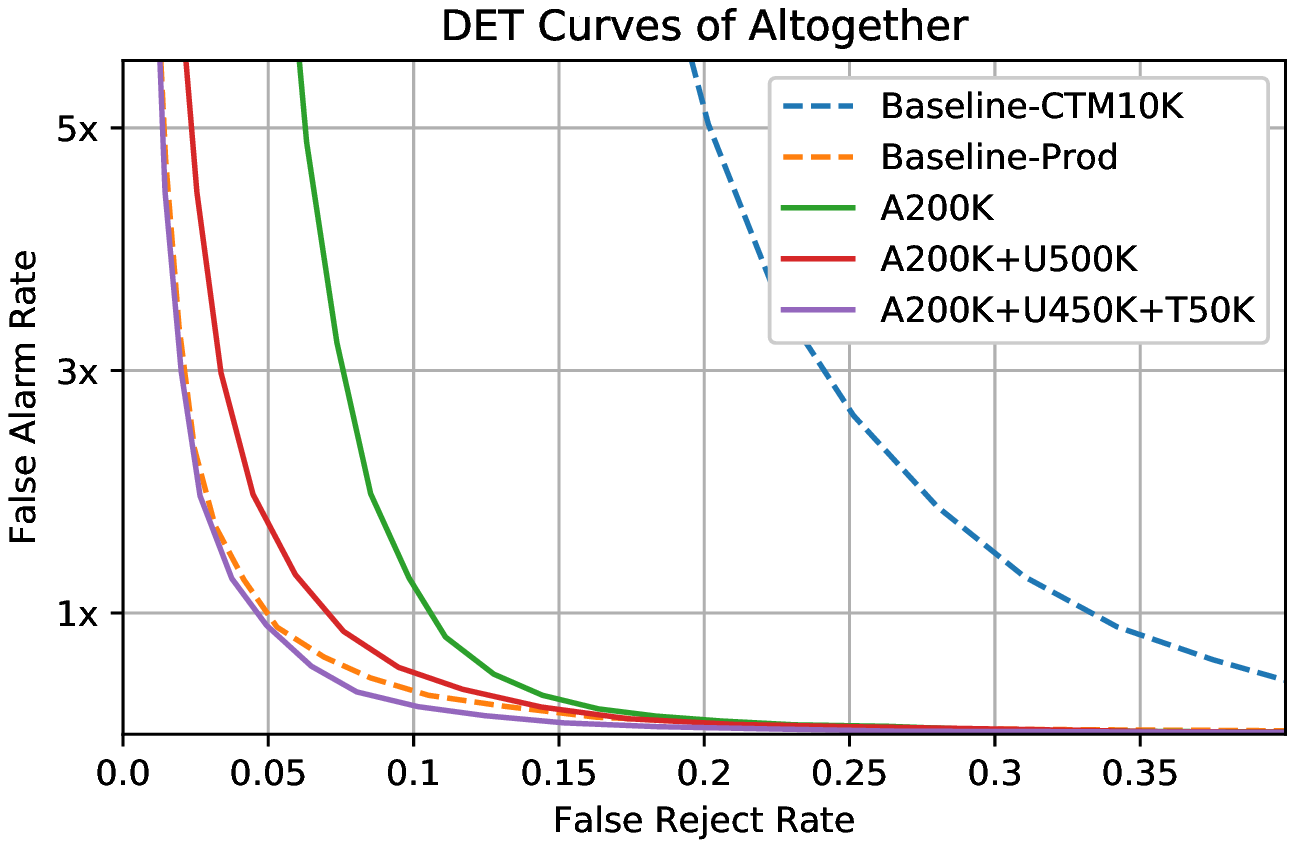}}
\centering
\caption{ DET curves with (a) multi-condition training sets of different sizes; (b) untranscribed and partial transcribed data; (c) multi-condition training and semi-supervised learning combined.}
\label{fig:results}
\end{figure*}
\vspace{-0.5em}
\subsection{Experimental setup}
%We study the effects of the proposed multi-condition training and semi-supervised learning strategies, respectively, and show how we are able bring the performance of a baseline model trained on 10 hrs CTM data (Baseline-CTM10K) to match the production-grade model trained with several hundreds hours of transcribed speech data collected from far-field devices in real use cases (Baseline-Prod) .%\footnote{Our experimental wake word system does not reflect the performance of the production Alexa system.}  

We use the GPU-based distributed trainer as described in \cite{Strom2015_Nemo} for DNN training. The models are trained using transfer-learning paradigm where the weights of the DNN are initialized by an ASR acoustic model of the same architecture and size trained with ASR senone targets. During training a multi-task learning framework similar to \cite{Panchi2016_Multitask, Sun2017_TDNN,Guo2018_Highway, Wu2018_Monophone, Raju2018_Augm} is used where the loss is a weighted sum of WW loss and ASR loss. The ASR branch regularizes the DNN training and is helpful especially when training dataset is small.

The evaluation metrics in this paper are false reject rate (FRR), which is 1 minus recall, and false alarm rate (FAR), which is a normalized number of false accepts. Absolute values of FAR in this paper have been anonymized for confidentiality reasons. The range of used FAR corresponds to the range normally used for production keyword spotting models. We report the two metrics in a detection-error-tradeoff (DET) curve as we tune the decision thresholds for each model. The lower the DET curve, the better the performance. 
\vspace{-0.5em}
\subsection{Results of multi-condition training}
%To study the performance of the multi-condition training using the data augmentation strategies described in Section \ref{sec:augm} , we 
Our data augmentation strategy defines the training sets to constitute conditions denoted as (CTM, CTM+R, CTM+N and CTM+RN), where CTM, R, N, RN denote the original close-talk microphone data, reverberation, noise corruption, reverberation followed by corruption, respectively. %, for better performance over broader range of applications from near-field to far-field. The components CTM, CTM+R, CTM+N and CTM+RN simulate the speech from  clean near-field, clean far-field, noisy near-field and noisy far-field, respectively.  
We sweep the size of the training set by increasing the portion of clean CTM data to as much as 10\% by repeated up-sampling, 
and mixing with equal amount of the three augmented datasets. The augmentation used in our experiments are detailed in Table \ref{tab:augmentation}%\footnote{The first dataset is actually 52K due to integer division. WLOG we call it 50K for clean notation.}.
\vspace{-0.5em}
\begin{table}[htp]
\centering
\caption{Details of the multi-condition training sets.}
\label{tab:augmentation}
\begin{tabular}{ccccc}
\hline
size & CTM & CTM+R&CTM+N&CTM+RN\\
\hline
50K&10K&14K&14K&14K\\
200K&20K&60K&60K&60K\\
350K&35K&105K&105K&105K\\
500K&50K&150K&150K&150K\\
\hline
\end{tabular}
\end{table}
\vspace{-0.5em}
%\begin{figure}[htp]
%\centering
%\includegraphics[width=0.4\textwidth]{figures/DET_MCT}
%\caption{ DET curves with multi-condition training sets of different sizes.}
%\label{fig:augm}
%\end{figure}

Fig. \ref{fig:results} (a) shows the DET curves of the augmented training sets (prefixed with ``A") of size 50K, 200K, 350K and 500K respectively,  compared to the two baselines: ``Baseline-CTM10K" (blue) and ``Baseline-Prod"(orange).  The DET curves show that the multi-condition training boosts the performance significantly. 
%However, the gain that multi-condition training could bring in is not unbounded. 
In our experiments 20x augmentation (A200K) achieves the maximum gain, which suggests that overwhelming augmentation may hurt the performance. %This is possibly due to the WW model starting to overfit the conditions present in the multi-condition dataset. 
There remains a gap between the best multi-condition DET curve (A200K) and the Baseline-Prod DET curve, indicating some mismatch exists between the simulated data and the real-world.
% This indicates that the multi-condition data augmentation can reproduce the real-world use cases to very significant extent (i.e. improve FAR by more than 20x at equivalent FRR). However, there remains a mismatch between the simulated multi-condition data and the real-world, so that the performance of a model trained with large amount of real-world data cannot be fully matched by using the multi-conditioning alone.
%
\vspace{-0.5em}
\subsection{Results  of semi-supervised learning}
%We study the performance of WW model trained with the untranscribed data using the semi-supervised learning (SSL) strategy described in section \ref{sec:ssl}. 
We use the the semi-supervised learning (SSL) strategy described in section \ref{sec:ssl} to construct two datasets (prefixed with ``U") of 500K and 1000K total utterances from the untranscribed in-house production speech corpus by employing a production-grade ASR model similar to \cite{Maas2017_Endpointing}.
%For the chosen WW,  we constructed two datasets of 500K and 1000K total utterances from the untranscribed in-house production speech corpus, respectively, and employed a production-grade ASR model similar to \cite{Maas2017_Endpointing} to obtain the automatic transcripts. 
Both datasets contain equal portion of utterances containing the WW and the confusable words as described in \ref{sec:confusable}. %The vocabulary of confusing words contains 57 words with phonetic Levenshtein distance from ``Amazon" of 1 and 2. The datasets are denoted as ``U500K" and ``U1000K" where ``U" means untranscribed.  
%\begin{figure}[tp]
%\centering
%\includegraphics[width=0.4\textwidth]{figures/DET_SSL}
%\caption{ DET curves with untranscribed and transcribed training data.}
%\label{fig:mining}
%\end{figure}

Fig. \ref{fig:results} (b) shows the DET curves of the WW model trained with those datasets. The untranscribed datasets also boost the WW detection performance significantly, and U1000K achieves more gain than U500K. %An interesting observation by comparing the DET curves of the untranscribed and augmented dataset is that they intersect, for instance the U500K and A200K curves intersect at FRR=0.08, which suggests that the untranscribed dataset performs better at lower FRR and the augmented dataset performs better at lower FAR.
Note that a gap still exists between the DET curves of U1000K and Baseline-Prod, which suggests that only untranscribed data is not enough to achieve production performance. We further transcribed 10\% of the 500K utterances to study the effect of human input and created the dataset called ``U450K+T50K" where "T" denotes transcribed. The model trained with this dataset outperforms the models with untranscribed-only data or the augmented-only data, and the DET curve is very close to the one of the Baseline-Prod. %Traditionally, training a production-grade model requires at least hundreds of hours of human-annotated WW-specific data. Comparatively, we are able to reduce the human effort significantly from hundreds of hours to approximately 50 hrs to achieve matching performance.  Additional, our data filtering method may also be used to pre-select data for human annotation. 
\subsection{Results of MCT and SSL combined }
Having observed that both the multi-condition training (MCT) and semi-supervised learning (SSL) approaches work better at different regions, we hypothesize that they supplement each other to achieve better performance if combined together. Therefore we take the best augmented dataset (A200K), and add to the untranscribed or the partially-transcribed dataset, denoted as A200K+U500K and A200K+U450K+T50K, respectively, and the result DET curves are shown in Fig. \ref{fig:results} (c). 
%\begin{figure}[tp]
%\centering
%\includegraphics[width=0.4\textwidth]{figures/DET_Altogether}
%\caption{ DET curves with multi-condition training and semi-supervised learning combined.}
%\label{fig:all}
%\end{figure}
We can observe that the A200K+U500K DET curve becomes much closer to the Baseline-Prod DET curve. 
%, indicating that this combination brings more performance gain. 
Moreover, with 50K utterances transcribed, the A200K+U450K+T50K DET curve is further improved and performs slightly better than Baseline-Prod model. This observation indicates that each approach (multi-condition training, semi-supervised learning, transcription) has own strength and limitation, and can supplement each other to achieve better than production-quality performance. %We further add more untranscribed data to create a dataset called ``A200K+T50K+U950K", while the gain from additional untranscribed data is observable but very little. Hence the gain from more untranscribed data may saturate at some point and it does not worth to add more and more untranscribed data, especially in consideration of the training cost. % For example at FRR=0.05, the False Alarm Rate is reduced by 4.5 times. %We further pulled more data from the untranscribed speech corpus using the same method, added to the A200K dataset, and denoted the dataset as ``A200K+P500K+N500K". The DET curve of this model was pushed further towards Baseline-Prod, for example at FRR=0.05, FAR is reduced by 6 times. However there is still a small gap between the Baseline-Prod and the ``A200K+P500K+N500K". .   
\vspace{-0.5em}
\section{Conclusion}
\label{sec:concl}
% conclusion

In this paper we proposed a series of methods based on multi-condition training and semi-supervised learning to model new wake word spotter using as little as 10 hours ($\sim$10 K) clean anechoic recordings.  By employing a stratified data augmentation to prepare 200K multi-condition data, added to 500K untranscribed utterances selected from a far-field speech corpus, together with transcribing 50K utterances, we are able to match a production-quality baseline. Comparatively, we saved 97\% (1-10/375) in data collection and 86\% (1-50/375) in transcription. 

Our work provides solutions to many different scenarios emerging during developing models for new WW, and the performance benchmarks (eg. @FRR=0.05): (1) If only the close-talk recordings are available, the multi-condition training brings us within 9x FAR from the Baseline-Prod; (2) if an untranscribed far-field speech corpus is available, the semi-supervised learning alone can bring us within 5x FAR; (3) if both augmented data and far-field corpus are available, combining the two can further bring us within 2x FAR; (4) if transcribed far-field data is present, or researchers have the ability to transcribe data, adding up to 50K transcribed utterances together with 10 times of untranscribed speech allow us to  build models that can match/outperform the production quality.

%Our work provides solutions to many different scenarios emerging during developing models for new WW:
%\begin{enumerate}
%\item If only the close-talk recordings are available, the multi-condition training method brings us within 9x FAR (@FRR=0.05) from the Baseline-Prod performance. In our experiments we find 20x augmented data to work the best in this situation; 
%\vspace{-0.2em}
%\item If an untranscribed far-field speech corpus is available, this alone can bring us within 5x FAR (@FRR=0.05) from the Baseline-Prod performance using the proposed semi-supervised learning method.% Generally speaking the more untranscribed data the better the performance, while training computation is another factor of consideration; 
%\vspace{-0.2em}
%\item If both augmented data and far-field corpus are available, combining the two can further bring us within 2x FAR (@FRR=0.05) from the Baseline-Prod performance;
%\vspace{-0.2em}
%\item If transcribed far-field data is present, or researchers have the ability to transcribe data, adding up to 50K transcribed utterances together with 10 times of untranscribed speech allow us to  build models that can match the production quality. 
%\end{enumerate} 
%
\bibliographystyle{IEEEbib}
\bibliography{wwbib}
\end{document}